\documentstyle[prl,aps,epsf]{revtex}
\begin{document}
\hfill{CPHT RR075.1102}
\vspace{1.cm}
\begin{center}
{\Large\bf
Genuine twist $3$ in exclusive electroproduction
of transversely polarized vector mesons}

\vspace{1cm}
{\sc I.V. Anikin}${}^{a,b}$ and
{\sc O.V.~Teryaev}${}^{a}$
\\[0.5cm]
\vspace*{0.1cm} ${}^a$
{\it
Bogoliubov Laboratory of Theoretical Physics, JINR, 141980 Dubna, Russia}
\\[0.2cm]
\vspace*{0.1cm} ${}^b$ {\it
CPhT, {\'E}cole Polytechnique, F-91128 Palaiseau, France\footnote{
  Unit{\'e} mixte C7644 du CNRS.}}
  \\[1.0cm]

\end{center}

%\maketitle

\begin{abstract}

\noindent
We present the detailed analysis of genuine twist-$3$ contributions
to the exclusive electroproduction amplitude of transversely polarized
vector mesons. Using the formalism based on the QCD factorization in
the momentum representation we calculated all the genuine twist-$3$ terms
and found the total expression of this amplitude at $1/Q$ level.
Generally speaking, these terms violate
standard factorization owing to
the existence of the infrared divergencies in
the amplitude of hard sub-processes, although the strongest
divergencies cancel due to the QCD
equations of motion.
We discuss the possible treatment of surviving divergencies.
\end{abstract}

\section{Introduction}

\noindent
Hard exclusive reactions provide important information
for unveiling the composite structure of hadrons.
Moreover, a self-consistent description of the hard exclusive reactions is
one of main goals of the QCD application. The perturbative theory
allows to implement the systematical calculations providing that, first,
it is possible to single out the amplitude of the short-distance
sub-process and, second, to prove the infrared finiteness of this amplitude.
Technically it corresponds to the factorization of large (hard) momenta
from small (soft) momenta domains. Besides, the factorized pieces of
amplitude depend on the dynamics that is typical at a given scale,
{\it i.e.} the amplitude becomes the convolution of hard and soft parts.
However, the electroproduction of transversely polarized vector mesons
\begin{eqnarray}
\label{elprod}
{\rm hadron}(p_1)+\gamma^*(q)\to\rho(p)+{\rm hadron}(p_2)\,\, ,
\end{eqnarray}
provides the well-known example of QCD factorization breaking.
Indeed, the factorization is valid
only in the
case of longitudinally polarized $\rho$-meson production
\cite{Col,Man1}, when
the factorization assert that the photon-parton
hard, perturbatively calculable, sub-processes
are parted from the nonperturbative matrix elements.

\noindent
Unfortunately, it is not so encouragingly simple for the case
of transversely polarized meson production.
Its description
is strongly complicated due to the existence of
infrared divergencies in amplitudes, breaking down
the factorization ( see e.g. \cite{Man2}, \cite{Rad}
and references therein).

The amplitude
of transverse vector meson production corresponds to the
contributions suppressed as $1/Q$
in comparison with the longitudinal vector meson case
\cite{Col}.
At the same time, the recent experiments show that the transverse
mesons production amplitudes provide a sensible contributions even at moderate
virtualities $Q^2$ \cite{Her}.
So, to describe these processes one should
take into account the   $1/Q$   terms.

\noindent
In Ref. \cite{Man1}, the analysis of twist-$2$
amplitudes for hard exclusive electroproduction of mesons in
terms of generalized parton distributions (GPD) was presented.
Later, the authors of \cite{Man2} have discussed the factorization
problems in the electroproduction of light vector mesons from
transversely polarized photons. They have taken into account the kinematical
twist-$3$ terms within the Wandzura--Wilczek (WW)
approximation.
Also, the helicity flip amplitude of transversely polarized vector
mesons production was considered within the same
approximation \cite{Kiv}.

Although the kinematical and dynamical (genuine) higher twists
contributions are, generally speaking, independent, there
are notable exceptions in  some kinematical regions.
In deep inelastic scattering at $x_B\to 1$ the
kinematical higher twist terms, described by Nachtmann
variable, lead to inconsistencies unless the genuine
higher twists are taken into account \cite{DGP}.
One cannot exclude, that the genuine higher twists
may cure, at least partially, the problem arising
from the treatment of the end-point regions in
hard electroproduction.

\noindent
Thus, in this paper our goal is
to study the role of genuine twist-$3$ contributions
in the factorization theorem breaking.
We adhere the approach based on
the momentum representation, the basic stages of which are expounded in
previous papers \cite{Ani1}-\cite{Ani3}.
The essence of this approach constitutes of
the generalization of Ellis--Furmanski--Petronzio
factorization scheme \cite{EFP}.
These authors have considered the twist-$4$ effects in the processes of
unpolarized deep inelastic scattering.
At the same time, our formalism is more closely following
the approach \cite{Efr}
developed by A.V. Efremov and one of the authors,
where the role of twist-$3$
terms in {\it polarized} deep inelastic scattering was studied in detail.

\noindent
In this paper we have computed the total expression for
transverse meson production
amplitude comprising the quark and gluon leading twist GPDs in nucleons
and the genuine twist-$3$ contributions to the $\rho$-meson wave
function.
As a cross-check, we have reproduced the
gluon contributions to the transverse meson production amplitude
in WW approximation obtained in \cite{Man2}.
We also have discussed the possible ways to treat
the infrared singularities and their partial cancellation.

\noindent
The structure of paper is as follows: in Sections II and III we
will introduce the kinematics and parameterization of matrix elements.
Also we will present the total amplitude of quark and gluon
distribution diagrams in Sections IV and V.
We make the conclusion in the last section.

\section{Kinematics}

\noindent
Let us start with the description of our kinematics:
the $p$ is momentum of transversely polarized $\rho$-meson and
its polarization vector is $e^T$; the momentum of virtual photon is denoted
by $q$ ($Q^2=-q^2$).
As the hard exclusive production kinematics
is similar to the DVCS kinematics, it is natural to use the
analogous notations (cf. \cite{Ani1}).
By making use of initial ($p_1$) and final ($p_2$) nucleon
momenta we construct the average momentum
$\overline{P}$ and transferred momentum $\Delta$:
\begin{eqnarray}
\label{PDel}
&&\overline{P}=\frac{p_2+p_1}{2}, \quad \Delta=p_2-p_1, \quad \Delta^2=t.
\end{eqnarray}
It is convenient to write up Sudakov decompositions for  all the
relevant particles.
We choose the light-cone basis composed by the physical vectors:
$\overline{P},\quad p $.
Such a choice is possible because in this paper
we assume that the initial hadron momentum $p_1$
and final hadron momentum $p_2$ are collinear, {\it i.e.} $\Delta^T\to 0$,
and $p_1^2=p_2^2=t=0$,
neglecting all the relevant higher twists contributions
arising from the nucleon matrix elements.
In addition, we neglect squares of meson masses.
Therefore the Sudakov decomposition up to kinematical twist-3
terms takes the form:
\begin{eqnarray}
\label{Sd}
&&\Delta=-2\xi\overline{P}, \quad e=e\cdot n\,p+e^T,
\quad n=\frac{\overline{P}}{p\cdot\overline{P}};
\nonumber\\
&&p=q-\Delta=p\cdot\overline{P}\,\tilde n;
\quad n\cdot\tilde n=\frac{1}{p\cdot\overline{P}}=\frac{4\xi}{Q^2},
\end{eqnarray}
where we introduced the normalized vectors $n$ and $\tilde n$.

\section{$\rho$-meson matrix elements and their properties}

\noindent
We introduce the parameterizations of
the $\rho$-meson--to--vacuum matrix elements needed for calculation
of amplitudes (cf. \cite{BB}).
Keeping all the terms up to the twist-$3$ order
and using the axial (light-like) gauge
\begin{eqnarray}
\label{g1}
n\cdot A=0,
\end{eqnarray}
these matrix elements can be written
in terms of the light-cone basis vectors (\ref{Sd}):
\begin{eqnarray}
\label{par1}
&&\langle \rho(p)|\bar\psi(0)\gamma_{\mu} \psi(z)|0\rangle
\stackrel{{\cal F}}{=}
\varphi_1(y)(e\cdot n)p_{\mu}+\varphi_3(y)e^{T}_{\mu},
\quad
\langle \rho(p)|
\bar\psi(0)\gamma_{\mu}
i\stackrel{\longleftrightarrow}
{\partial^T_{\rho}} \psi(z)|0 \rangle
\stackrel{{\cal F}}{=}
\varphi_1^T(y)p_{\mu} e^T_{\rho},
\\
\label{par1.1}
&&\langle \rho(p)|
\bar\psi(0)\gamma_5\gamma_{\mu} \psi(z) |0\rangle
\stackrel{{\cal F}}{=}
i\varphi_A(y)\varepsilon_{\rho\alpha\beta\delta}
e^{T}_{\alpha}p_{\beta}n_{\delta},
\quad
\langle \rho(p)| \bar\psi(0)\gamma_5\gamma_{\mu}
i\stackrel{\longleftrightarrow}
{\partial^T_{\rho}} \psi(z) |0\rangle
\stackrel{{\cal F}}{=}
i\varphi_A^T (y) p_{\mu}\varepsilon_{\rho\alpha\beta\delta}
e^{T}_{\alpha}p_{\beta}n_{\delta},
\\
\label{par1.2}
&&\langle \rho(p)|
\bar\psi(0)\gamma_{\mu}g A_{\rho}^T(z_2) \psi(z_1) |0\rangle
\stackrel{{\cal F}}{=}
\Phi(y_1,y_2)p_{\mu} e^{T}_{\rho},
\nonumber\\
&&\langle \rho(p)|
\bar\psi(0)\gamma_5\gamma_{\mu} g A_{\rho}^T(z_2) \psi(z_1) |0\rangle
\stackrel{{\cal F}}{=}
i J(y_1,y_2)p_{\mu}\varepsilon_{\rho\alpha\beta\delta}
e^{T}_{\alpha}p_{\beta}n_{\delta},
\nonumber\\
\label{cov}
&&\langle \rho(p)|
\bar\psi(0)\gamma_{\mu}i\stackrel{\longleftrightarrow}
{D^T_{\rho}}(z_2) \psi(z_1) |0\rangle
\stackrel{{\cal F}}{=}
\tilde \Phi(y_1,y_2)p_{\mu} e^{T}_{\rho},
\nonumber\\
&&\langle \rho(p)|
\bar\psi(0)\gamma_5\gamma_{\mu} i\stackrel{\longleftrightarrow}
{D^T_{\rho}}(z_2) \psi(z_1) |0\rangle
\stackrel{{\cal F}}{=}
i \tilde J(y_1,y_2)p_{\mu}\varepsilon_{\rho\alpha\beta\delta}
e^{T}_{\alpha}p_{\beta}n_{\delta}
 \nonumber \\
&&\tilde\Phi(y_1,y_2)=\frac{1}{2}\biggl(
\varphi_1^T(y_1)+\varphi_1^T(y_2)\biggr)\delta(y_1-y_2)+\Phi(y_1,y_2),
\nonumber\\
&&\tilde J(y_1,y_2)=\frac{1}{2}\biggl(
\varphi_A^T(y_1)+\varphi_A^T(y_2)\biggr)\delta(y_1-y_2)+J(y_1,y_2)
\end{eqnarray}
here $\stackrel{\longleftrightarrow}
{\partial_{\rho}}=\frac{1}{2}(\stackrel{\longrightarrow}
{\partial_{\rho}}-\stackrel{\longleftarrow}{\partial_{\rho}})$
is the standard antisymmetric derivative and $\stackrel{{\cal F}}{=}$
denotes the Fourier transformation
with measure ($z_i=\lambda_i n$):
\begin{eqnarray}
dy \,e^{ -iy\,pz}
&&\quad {\rm for\,\, quark\,\, correlators},
\nonumber\\
dy_1 \,dy_2 \,e^{ -iy_1\,pz_1
-i(y_2 - y_1)\,pz_2 }
&&\quad {\rm for\,\, quark-gluon\,\, correlators}.
\end{eqnarray}

\noindent
Note that the function $\varphi_1$ corresponds to
the twist-$2$; the functions $\varphi_1^T$,
$\varphi_A^T$ correspond to the WW twist-$3$,
functions $\Phi$ and $J$ -- to the genuine (dynamical) twist-$3$,
while functions $\varphi_3$, $\varphi_A, \tilde \Phi, \tilde J$
contain both.

\noindent
In (\ref{par1})--(\ref{par1.2}) the functions  $\varphi_1$,
 $\varphi_3$ and $\varphi_A$ parameterizing the two-particle correlators
obey the following symmetry properties:
\begin{eqnarray}
\label{sym1}
\varphi_1(y)=\varphi_1(1-y), \quad  \varphi_3(y)=\varphi_3(1-y), \quad
\varphi_A(y)=-\varphi_A(1-y).
\end{eqnarray}
At the same time, the symmetry properties of the
functions $\Phi$ and $J$ parameterizing the three-particle
correlators are:
\begin{eqnarray}
\label{sym2}
\Phi(y_1,y_2)=\Phi(1-y_2,1-y_1), \quad J(y_1,y_2)=-J(1-y_2,1-y_1).
\end{eqnarray}
The relations (\ref{sym2}) represent the particular case of the relations
for the functions parameterizing the analogous matrix elements
in the production of pion pair with the arbitrary angular momentum
$j$\cite{Ani2},
while the case under consideration corresponds, obviously, to $j=1$.
In the case of arbitrary $j$ the two-body
functions depend on the extra parameter
$\xi=(p_{\pi}-p^{\prime}_{\pi})\cdot n$
characterizing the skewedness
and they have the following properties
\footnote{Here we correct the misprints in
\cite{Ani2}, formula (48). Notations:
$\Phi^{\pi\pi}$ and $J^{\pi\pi}$ correspond to
$\tilde B$ and $\tilde D$, respectively}:
\begin{eqnarray}
\label{sym22}
\Phi^{\pi\pi}(y_1,y_2;\xi)=\Phi^{\pi\pi}(1-y_2,1-y_1;-\xi),
\quad J^{\pi\pi}(y_1,y_2;\xi)=-J^{\pi\pi}(1-y_2,1-y_1;-\xi).
\end{eqnarray}

\noindent
For the sake of comparison, we show the relations between our
parameterizing functions introduced in (\ref{par1}) and functions
used in \cite{BB}:
\begin{eqnarray}
\label{relBB}
\varphi_1(y)\Leftrightarrow
f_{\rho}m_{\rho}\phi_{\parallel}(y) ,
\quad
\varphi_3(y)\Leftrightarrow
f_{\rho}m_{\rho} g^{(v)}(y),
\quad
\varphi_A(y)\Leftrightarrow
-\frac{1}{4}f_{\rho}m_{\rho} \frac{\partial g^{(a)}(y)}{\partial y}.
\end{eqnarray}
Let us remind that the functions $\varphi_3(y)$ ($g^{(v)}(y)$) and
$\varphi_A(y)$ ($g^{(a)}(y)$) can be expressed through
the $\varphi_1(y)$ ($\phi_{\parallel}(y)$) owing to
the WW-type relations \cite{Ani2} (\cite{BB}).
\noindent
The WW approximations take the
especially simple form in terms of functions \cite{Ani2}
\begin{eqnarray}
\label{com_pl}
\varphi_{\pm}(y)=\varphi_3(y)\pm \varphi_A(y).
\end{eqnarray}
WW relations for $\rho$ meson represent the particular case of these
relations for
pion pair with arbitrary angular momentum \cite{Ani2} and take the form:
\begin{eqnarray}
\label{sol_WW}
\varphi_+^{WW}(x)= -
\int\limits_{x}^{1}\frac{dy}{y}
\varphi_1(y) ,
\quad
\varphi_-^{WW}(x)=
\int\limits_{0}^{x}\frac{dy}{y-1}
\varphi_1(y).
\end{eqnarray}

\noindent
Let us turn to the integral relations
arising from the QCD equations of motion, which may be derived
closely following \cite{Ani1} and \cite{Ani2}:
\begin{eqnarray}
\label{em_rho}
&&\int\limits_{0}^{1} dy \left( \tilde \Phi^{(S)}(x,y)
- \tilde J^{(A)}(x,y) \right) =
\left(x-\frac{1}{2}\right)
\varphi_3(x)+\frac{1}{2} \varphi_A(x),
\nonumber\\
&&\int\limits_{0}^{1} dy \left( \tilde \Phi^{(A)}(x,y)
- \tilde J^{(S)}(x,y) \right) =
-\left(x-\frac{1}{2}\right)
\varphi_A(x)-\frac{1}{2} \varphi_3(x),
\end{eqnarray}
where symmetric and anti-symmetric functions,
whose  appearance is crucial for the compatibility with the
symmetry properties (\ref{sym1}, \ref{sym2}), are defined as
$(f=\tilde \Phi , \, \tilde J)$:
\begin{eqnarray}
\label{saf}
&& f^{(S,A)}(x,y)=\frac{1}{2}
\left( f(x,y) \pm f(y,x) \right).
\end{eqnarray}
Moreover, the symmetry properties make
two equations equivalent to the single one, which may be rewritten
in the simple form (c.f. \cite{Efr}):
\begin{eqnarray}
\label{em2}
\int\limits_{0}^{1} dy_2 \tilde{\cal F}_{-}(y_1,y_2) =
\left(1-y_1\right) \varphi_{-}(y_1).
\end{eqnarray}
Here and below we use the obvious similar notations for
quark-gluon correlators:
\begin{eqnarray}
\label{Comb1}
{\cal F}_{\pm}(y_1,y_2)= J(y_1,y_2) \pm \Phi(y_1,y_2),\,\,\
\tilde {\cal F}_{\pm}(y_1,y_2)=\tilde J(y_1,y_2) \pm \tilde \Phi(y_1,y_2).
\end{eqnarray}
As we will see in the next section,
the QCD equations of motion play the important role
in the cancellation of some leading infrared divergencies
of the transverse $\rho$-meson
production amplitudes.

\section{Amplitude of quark-photon scattering}

\noindent
At first we dwell on the computation of
the quark contributions to the production amplitude
\footnote{Recently we computed this amplitude using another
parameterization \cite{Ani4} }.
As we would like to concentrate on the twist-$3$ effects in
$\rho$-meson blob in Fig.1,  it is sufficient to
keep the twist-$2$ GPD
in the parameterization of quark matrix elements
over nucleon states. Hence, taking into account (\ref{PDel}) and (\ref{Sd})
the parameterization acquires the following form
\begin{eqnarray}
\label{parhad-q}
\langle N(p_2)|\bar\psi(0)\gamma_{\mu} \psi(\tilde z)
 | N(p_1)\rangle
\stackrel{{\cal F}}{=}H(x)\overline{U}(p_2)\gamma_{\mu}U(p_1)
=\sqrt{1-\xi^2}H(x)\overline{P}_{\mu}
\end{eqnarray}
where, as it was above,
$\stackrel{{\cal F}}{=}$ denotes the Fourier transformation,
except  that
the measure now is ($\tilde z=\lambda \tilde n$)
\begin{eqnarray}
\label{mea}
dx e^{ -i( x \overline{P}+\Delta/2)\tilde z }.
\end{eqnarray}
Further, using the parameterization (\ref{par1})--(\ref{par1.2}),
(\ref{parhad-q}) and calculating the traces,
the amplitude given by the simplest Feynman diagrams, Figs. 2a and 2b,
reads (cf. \cite{Man1}):
\begin{eqnarray}
\label{qdsim}
{\cal A}_{1,\,\mu}^{(q),\,\gamma_T^*\to\rho_T}=
8\sqrt{1-\xi^2}\,\frac{C_F}{N_c}\,\frac{e_{\mu}^T}{Q^2}
\,
\int_{-1}^{1}dx H(x)\biggl[
\frac{1}{x-\xi+i\epsilon}-\frac{1}{x+\xi-i\epsilon}
\biggr] {\cal S}_2^{(q)},
\end{eqnarray}
where
\begin{eqnarray}
\label{S1}
{\cal S}_2^{(q)}=\int_{0}^{1}\frac{dy}{y} \varphi_{+}(y).
\end{eqnarray}
Note that the double poles in $x$ cancel due to the use of
the gluon propagator in the axial gauge.

\noindent
It is convenient to organize genuine twist 3 diagrams according to the
insertions of extra gluon to the diagrams of Fig.2.
Summing the diagrams
of Figs. 3 and 4, we finally obtain the
expression for the quark distribution amplitude that includes
all the WW and genuine twist-$3$ contributions:
\begin{eqnarray}
\label{qdtw3}
{\cal A}_{2,\,\mu}^{(q),\,\gamma_T^*\to\rho_T}=
8\sqrt{1-\xi^2}\,\frac{C_F}{N_c^2-1}\,\frac{e_{\mu}^T}{Q^2}
\,\Biggl\{
{\cal H}_1\times {\cal I}_1^{(q)} +
{\cal H}_2\times {\cal I}_2^{(q)}
\,\Biggr\},
\end{eqnarray}
where
\begin{eqnarray}
\label{H}
{\cal H}_1=\xi\int_{-1}^{1}dx H(x)\biggl[
\frac{1}{(x+\xi-i\epsilon)^2}+\frac{1}{(x-\xi+i\epsilon)^2}
\biggr] ,
\quad
{\cal H}_2=
\int_{-1}^{1}dx H(x)\biggl[
\frac{1}{x-\xi+i\epsilon}-\frac{1}{x+\xi-i\epsilon}
\biggr]
\end{eqnarray}
and
\begin{eqnarray}
\label{I1q}
{\cal I}_1^{(q)}=\int_{0}^{1} dy_1\,dy_2
\Biggl\{
\frac{4C_F \tilde {\cal F}_{-}(y_1,y_2)}{(1-y_1)^2}
+
\frac{C_A {\cal F}_{-}(y_1,y_2)}{(1-y_2)(1-y_1)}
+\frac{2C_F\tilde {\cal F}_{-}(y_1,y_2)
-C_A{\cal F}_{-}(y_1,y_2)}{y_1(1+y_1-y_2)}
\Biggr\};
\end{eqnarray}
\begin{eqnarray}
\label{I2q}
{\cal I}_2^{(q)}=\int_{0}^{1} dy_1\,dy_2
\Biggl\{
\frac{ C_F\tilde {\cal F}_{-}(y_1,y_2)}{(1-y_2)(1-y_1)}
+
\frac{2C_F\tilde {\cal F}_{-}(y_1,y_2)
-C_A{\cal F}_{-}(y_1,y_2) }{(1+y_1-y_2)(1-y_2)}
\Biggr\}.
\end{eqnarray}
Note that the entire result is expressed in terms of
$"-"$ combinations of twist-3 functions
(which may be transformed
to $"+"$ ones by making use of symmetry properties (\ref{sym2})).
Such combinations of axial and vector twist-3
matrix elements are known to appear also
in inclusive case \cite{Efr,ET85,EKT}.

\noindent
The ${\cal H}_1$- and ${\cal I}_1^{(q)}$-structure integrals
possess the poles of second order. This fact leads to the vulnerability
of factorization theorem.
However, one can see that the first term in (\ref{I1q}) may be reduced
to the two-particle function $\varphi_-(y_1)$ by making use of
the QCD equations of motion (\ref{em2}).
As a result, the power of infrared divergence is reduced due to the factor
$1-y_1$ in its r.h.s.
Moreover, in the case of the gauge different from the axial one
(Feynman gauge, in particular)
the emerging double pole in $x$ in (\ref{qdsim}) is cancelled
by the corresponding additional contribution to (\ref{qdtw3}) (together with
the dependence on the gauge fixing parameter) provided the
equation of motion are taken into account.
This is quite natural:
while the longitudinal amplitude is gauge invariant by itself,
the gauge invariant set of diagrams for the transverse amplitude
contains also quark gluon diagrams, related to the quark ones by the
equations of motion.

\noindent
Still, the remaining single poles would lead to infrared divergencies
unless functions
$\Phi, \tilde\Phi$,  $J, \tilde J$ vanish at $y_i\to 1$ or $y_i\to 0$
at least as the first power of $(1-y_i)$ or $y_i$.
This behaviour is quite reasonable for genuine twist term.
However, as the corresponding integral (with a colour factor $C_A$)
becomes finite providing they have the corresponding behaviour
at the edge points, it cannot be used to cancel the infrared divergence of WW
contribution in the term with the factor $C_F$.

\noindent
To be more specific,
let us calculate this WW contribution explicitly.
Within the WW approximation where all
the quark-gluon parameterizing
functions $J(y_1,y_2)$ and $\Phi(y_1,y_2)$ (see, (\ref{par1.2}))
are equal to zero, the integrals at the structures
${\cal H}_1$ and ${\cal H}_2$  in the sum of (\ref{qdsim}) and
(\ref{qdtw3}) take the following forms:
\begin{eqnarray}
\label{IqWW}
&& {\cal K}_1^{(q)}=
{\cal I}_1^{(q),WW}=-\frac{1}{2}\int_{0}^{1} dy \,
\varphi_{-}(y) \biggl( 1+\frac{2}{y} \biggr)
\nonumber\\
&&{\cal K}_2^{(q)}={\cal S}_2^{(q)}+
{\cal I}_2^{(q),WW}=\int_{0}^{1} dy
\biggl(\varphi_{+}(y)+\frac{\varphi_{+}(y)+\varphi_{-}(y)}{2y}\biggr)
\end{eqnarray}

One can see from (\ref{sol_WW}) that the simple
pole in ${\cal K}_1^{(q)}$ of
(\ref{IqWW}) is cancelled by the zero boundary condition for the
function $\varphi_{-}(y)$~: $\varphi_{-}(0)=0$.
This is quite similar to the cancellation providing the
finiteness of the two pions production amplitude in the collision
of real and virtual photons \cite{Ani2}.
This makes finite the coefficient of
the potentially dangerous (due to the double pole in $x$) integral ${\cal H}_1$.
At the same time, there is no such effect for ${\cal K}_2^{(q)}$.

\section{Amplitude of gluon-photon scattering}

\noindent
In this section we turn to
the calculation of gluon distribution amplitude.
Since we will deal with the two-gluon GPD,
it is natural to choose the gauge-fixing condition
for the gluons in the form:
\begin{eqnarray}
\label{g2}
\tilde n\cdot A=0,
\end{eqnarray}
where $\tilde n$-vector  (\ref{Sd}) corresponds to the "minus"
component of the light-cone basis where "plus" component is provided
by nucleon (average) momentum.
In this case, the gluon fields $A_{\mu}$ can be expressed
through the gauge-invariant field-strength tensor $G_{\mu\nu}$.
Consequently, one can parameterize the nucleon matrix elements of
two gluon fields in terms of the gauge-invariant gluon distribution:
\begin{eqnarray}
\label{parhad-g}
\langle N(p_2)|
A_{\alpha}^{a}(0)A_{\beta}^{b}(\tilde z)
| N(p_1)\rangle
\stackrel{{\cal F}}{=}
\frac{\delta^{ab}}{N_c^2-1}
\biggl( g_{\alpha\beta}-
\overline{P}_{\alpha}\tilde n_{\beta}-
\overline{P}_{\beta}\tilde n_{\alpha}
\biggr)
\frac{G(x)}{(x+\xi-i\epsilon)(x-\xi+i\epsilon)}.
\end{eqnarray}

\noindent
Note that the gauge condition we use here is different from the
one used for parameterization of vector meson matrix elements
(see, (\ref{par1})-(\ref{par1.2}));
so the question arises, is it possible to use these expressions here.
Therefore, let us discuss this problem in more detail.

\noindent
In principle, owing to the Lorentz and gauge invariance, our physical
amplitudes should be independent on the explicit choice form of
$n$ and $\tilde n$.
In the general case, the vectors $n$ and $\tilde n$
can be chosen in an arbitrary way \cite{Ani2}.

\noindent
Now we would like to
make the comparative analysis of two gauges: (\ref{g1}) and (\ref{g2}), used
for $\rho$-meson matrix elements.
In the case of gauge (\ref{g1}) that was also used for the quark-gluon
scattering calculation,
the vector $n$ in (\ref{par1})-(\ref{par1.2})
play the double role. Namely, it fixes the gauge and
defines the longitudinal momentum fraction $x=kn$
carried by the active quark with the momentum $k$ in $\rho$-meson.
At the same when the gauge (\ref{g2}) is adopted,
these two roles are distributed
between the vectors $n$ and $\tilde n$. Namely, $n$ defines the
longitudinal momentum fraction whereas $\tilde n$ defines
the gauge-fixing condition.
Of course, in this case the parameterization of relevant
$\rho$-meson matrix elements should, generally speaking,
look more complicated due to
the added $\tilde n$-terms.
However, for "physical" $n, \tilde n$ (\ref{Sd}), one may explicitly check
that the parameterization (\ref{par1.1})-(\ref{cov}) is still self-consistent.
This is actually due to the twist-$3$ approximation, while there
is no hope to describe twist-$4$ terms in such a simple way.

\noindent
In the gauge (\ref{g1}) only
the transverse (physical) gluons exist.
In contrast, in the gauge (\ref{g2})
the transverse gluon fields are constructed as
\begin{eqnarray}
\label{g-field2}
A_{\rho}^T=A_{\rho}-p_{\rho} n\cdot A,
\end{eqnarray}
where the second term in (\ref{g-field2}) corresponds to the twist-$2$
contribution.
As a result, the use of gauge (\ref{g2}) leads to the appearance of twist-2
functions in the parameterization of three-particles
$\rho$-meson matrix elements, for instance
\begin{eqnarray}
\label{par1.22}
&&\langle \rho(p)|
\bar\psi(0)\gamma_{\mu}g A_{\rho}(z_2) \psi(z_1) |0\rangle
\stackrel{{\cal F}}{=}
\Phi_0(y_1,y_2)(e\cdot n)p_{\mu} p_{\rho} +
\Phi(y_1,y_2)p_{\mu} e^{T}_{\rho}.
\end{eqnarray}
The new twist-$2$ terms, proportional to $n\cdot A$,
should be absorbed to the standard P-ordered exponent (the gauge link) of
the matrix elements of leading twist-$2$ quark operators \cite{Qiu}.

\noindent
Thus, with the help of parameterization
(\ref{par1})--(\ref{par1.2}) and (\ref{parhad-g})),
we sum the amplitudes given by simplest diagrams of Figs. 2c
and 2d, with the amplitudes corresponding to diagrams of Figs. 5 and 6.
The last sort of amplitudes includes  both the kinematical and
dynamical (genuine) twist-$3$ in $\rho$-meson-to-vacuum matrix elements.
We derive the following expressions:
the simplest diagrams contributions reads
\begin{eqnarray}
\label{sgd}
{\cal A}_{1,\mu}^{(g),\,\gamma_T^*\to\rho_T}=
8\,\frac{C_F}{N_c^2-1}\,\frac{e_{\mu}^T}{Q^2}
&&\Biggl\{
\int_{-1}^{1}dx G(x)
\biggl[
\frac{1}{(x+\xi-i\epsilon)^2}+\frac{1}{(x-\xi+i\epsilon)^2}
\biggr]
{\cal S}_1^{(g)} +
\Biggr.
\nonumber\\
\Biggl.
&&\frac{1}{\xi}\int_{-1}^{1}dx G(x)
\biggl[
\frac{1}{x-\xi+i\epsilon}-\frac{1}{x+\xi-i\epsilon}
\biggr]
{\cal S}_2^{(g)}
\Biggr\},
\end{eqnarray}
where
\begin{eqnarray}
\label{Sg}
{\cal S}_1^{(g)}=
\int_{0}^{1} \frac{dy}{y}\,
\biggl( \frac{1}{2}\varphi_+(y)-\frac{3}{2}\varphi_-(y)\biggr),
\quad
{\cal S}_2^{(g)}=
\int_{0}^{1} dy\,
\frac{\varphi_+(y)+\varphi_-(y)}{y}
\end{eqnarray}
and the genuine twist-$3$ contributions take the form
\begin{eqnarray}
\label{gdtw3}
{\cal A}_{2,\mu}^{(q),\,\gamma_T^*\to\rho_T}=
8\,\frac{C_F N_c}{(N_c^2-1)^2}\,\frac{e_{\mu}^T}{Q^2}
\Biggl\{ {\cal G}_1 \times {\cal I}_1^{(g)} + {\cal G}_2 \times {\cal I}_2^{(g)}
\Biggr\},
\end{eqnarray}
where
\begin{eqnarray}
\label{G}
{\cal G}_1=\int_{-1}^{1}dx G(x)\biggl[
\frac{1}{(x+\xi-i\epsilon)^2}+\frac{1}{(x-\xi+i\epsilon)^2}
\biggr] ,
\quad
{\cal G}_2=\frac{1}{\xi}
\int_{-1}^{1}dx G(x)\biggl[
\frac{1}{x-\xi+i\epsilon}-\frac{1}{x+\xi-i\epsilon}
\biggr]
\end{eqnarray}
and
\begin{eqnarray}
\label{I1g}
{\cal I}_1^{(g)}=\int_{0}^{1} dy_1\,dy_2
\frac{C_A {\cal F}_-(y_1,y_2)}{(1-y_2)(1-y_1)},
\quad
{\cal I}_2^{(g)}=\int_{0}^{1} dy_1\,dy_2
\frac{ C_F\tilde{\cal F}_-(y_1,y_2)
+C_A {\cal F}_-(y_1,y_2)}{(1-y_2)(y_1-1)}.
\end{eqnarray}
Note that the integral ${\cal I}_1^{(g)}$ is essentially non-Abelian
and contains only genuine twist-3 contribution.
Consequently, the assumption on the linear decrease of ${\cal F}_-(y_1,y_2)$
leads to its finiteness.
At the same time, the integral ${\cal I}_2^{(g)}$ becomes divergent
due to WW term with the coefficient $C_F$.

\noindent
To study these effects and to
reproduce the results obtained in \cite{Man2} we address WW
approximation. As in the quark case, this approximation implies that all
the quark-gluon parameterizing
functions $J(y_1,y_2)$ and $\Phi(y_1,y_2)$ (see, (\ref{par1.2}))
are equal to zero. Using the results of \cite{Ani2} after some
computations we derive the following expression:
\begin{eqnarray}
\label{gdWW}
{\cal A}_{WW,\mu}^{(g),\,\gamma_T^*\to\rho_T}=
\frac{4}{N_c}\,\frac{e_{\mu}^T}{Q^2}
\,\Biggl\{ {\cal G}_1\times {\cal N}_1^{(g)} +
{\cal G}_2\times {\cal N}_2^{(g)}
\,\Biggr\},
\end{eqnarray}
where
\begin{eqnarray}
\label{I1WW}
&&{\cal N}_1^{(g)}={\cal S}_1^{(g)}=
\int_{0}^{1}\frac{dy}{y}
\Biggl( \frac{1}{2}\varphi_+(y)-\frac{3}{2}\varphi_-(y)\Biggr),
\nonumber\\
&&{\cal N}_2^{(g)}={\cal S}_2^{(g)}+
{\cal I}_2^{(g),\,WW}=\int_{0}^{1}\frac{dy}{y}
\Biggl( \frac{3}{2}\varphi_+(y)-\frac{1}{2}\varphi_-(y)\Biggr).
\end{eqnarray}
If we take into account (\ref{relBB}) then the integrals (\ref{I1WW})
will be rewritten in the forms which completely coincide with \cite{Man2}
\footnote{The definition of function $\Phi_{\parallel}$ can be found
in \cite{BB}.}:
\begin{eqnarray}
\label{I1BB}
&&{\cal N}_1^{(g)}=\int_{0}^{1}\frac{dy}{y}
\Biggl( 2 g^{(v)}(y)+\frac{g^{(a)}(y)}{2y(1-y)}\Biggr) ,
\nonumber\\
&&{\cal N}_2^{(g)}=\int_{0}^{1}\frac{dy}{y}
\Biggl( 4g^{(v)}(y)-2\frac{\Phi_{\parallel}(y)}{y}+
\frac{g^{(a)}(y)}{2y(1-y)} \Biggr).
\end{eqnarray}
%As we see the amplitudes (\ref{sgd}) and (\ref{gdtw3})
%is the product of two typical
%integrals ${\cal G}_i$ and ${\cal I}_i^{(g)}$.
%One of them ${\cal G}_i$
%is related to the gluon GPD and another one
%${\cal I}_i^{(g)}$ to the meson distribution amplitude.
Similar to the quark distribution case,
the  integrals (\ref{I1WW}) (and, consequently, (\ref{I1BB})) are not free
from  the infrared divergencies
owing to non-zero boundary values \cite{Man2,BB}.
%So, the main result of this paragraph is the following:
%we have derived within our approach the new contributions of the genuine twist $3$ and
%reproduced as a cross-check the known results.

\section{Discussion and Conclusions}

\noindent
%In the following we briefly discuss the obtained results and
%qualitatively suggest how we can regularize the infrared divergencies
%in the production amplitudes.
Let us stress that we calculated the full set of
genuine twist 3 diagrams of transverse
vector meson electroproduction. We also present, for the first time,
the WW contribution proportional to quark GPD and reproduce the WW
contribution due to the gluon GPD calculated earlier by the different method.
As a result, we observed the cancellation
of the few types of the leading infrared divergencies:

i) The double poles in $x$ of the quark contribution (Fig 2a,b)
in the non-axial gauge
cancel with the contributions of quark-gluon diagrams (Fig 3,4),
provided the QCD equations of motion are taken into account.

ii) Making use of the equation of motion allows also to eliminate
the double poles in $y$ of the same quark-gluon diagrams, surviving in the
axial gauge, which are reduced to the single poles.

iii) The single pole in $y$ of the WW contribution proportional to
the integral ${\cal H}_1$ of quark GPD
cancels between the vector ($\phi_3$) and axial ($\phi_A$) distributions.

\noindent
Let us start the analysis of potential surviving divergencies
from the double poles in $x$.
Consider first
%The gluonic contribution is known to be dominant at large energies, when
%the amplitude is predominantly imaginary.
the {\it imaginary} parts of $ {\cal H}_1,{\cal G}_1$
(providing in the gluon case the dominant contribution at large energies),
which are easily calculated and is represented
by the derivatives $
\frac{\partial}{\partial x}H(x,\xi)|_{x=\xi},
\frac{\partial}{\partial x}G(x,\xi)|_{x=\xi}$ in the
transition point $x=\xi$ from DGLAP to ERBL region. One may worry,
that for the quark case this derivative
is not continuous just at that transition point
due to the existence of so-called Polyakov-Weiss term
in the two-component model of GPD \cite{PW},
which is non-zero only for $|x|<\xi$. However, this term does not
give any contribution to the imaginary part of amplitude(like any
meson exchange term), and should not
therefore be taken into account when derivative is calculated, so that
it should be understood as
$\frac{\partial}{\partial x}H(x,\xi)|_{x=\xi+\epsilon}$.
Another way of PW term treatment \cite{Radon} is the consideration of its
most general form, occupying the whole region $|x|<1$,
and being therefore smooth at $x= \pm \xi$.
It may be reduced to the original PW form
by the sort of "gauge transformation" \cite{Radon}, generating the
irregularities both in PW term and another component of GPD, so
that they are cancelled in their sum.
The possible discontinuity of derivative in the gluon case
may be treated by adding the small mass to the quark propagators
\cite{ET85}, resulting in the similar expression
$\frac{\partial}{\partial x}H(x,\xi)|_{x=\xi+\epsilon}$.
Anyway, the imaginary parts of $ {\cal H}_1,{\cal G}_1 $ are
well defined, while the
real parts may be restored by making use of the dispersion relations \cite{disp}.

\noindent
Let us now discuss the surviving divergencies arising from the integration
over $y$.
The truly non-Abelian higher twist contributions ($C_A$ terms)
may be assumed finite, as it is sufficient to have the
functions $\Phi(y_1,y_2), J(y_1,y_2)$
which are going to zero when $y_{1,2}, \bar y_{1,2} \to 0$.
The counterpart of this assumption is, however, the impossibility
to cancel the divergencies of WW terms, which contain another colour
factor $C_F$.

\noindent
The only real danger is therefore coming from the WW terms, which
have the non-zero values at of $\phi_3, \phi_A$ at $y=0,1$.
Even if they are assumed to be zero
at some reference point $Q_0^2$, this property would be (rather slowly)
broken by the QCD
evolution.
%We cannot exclude the cancellation of this terms by the genuine twist
%ones, although this would also require further investigation.

\noindent
A visible way to regularize these singularities is
to keep the transverse momentum $k_i^T$ in loop integrations, which is in fact
required for the quantitative description of the longitudinal
amplitude as well \cite{Marc}.
At large $Q^2$ the behavior of structure integral is
determined by the region where, for instance, $1-y$ is of the same order
as $\langle k^2_T\rangle/Q^2$.
%( the $\langle k^2_T\rangle$ is the
%average value of "transversality" determined phenomenologically).
In order  to estimate the relevant contribution,
staying in the collinear approximation,
one may implement the corresponding infrared cutoff in the integration over $y$.
As a result, the transverse amplitude is logarithmically enhanced, so that
this non-factorizable contribution defines the asymptotic behaviour of
the ratio of transverse and longitudinal amplitudes:
\begin{equation}
\frac{|{\cal A}_{\mu}^{(q),\,\gamma_T^*\to\rho_T}|}
{|{\cal A}_{\mu}^{(q),\,\gamma_L^*\to\rho_L}|} \sim \frac{m_\rho ln Q}{Q},
\end{equation}
where power suppression comes from the standard kinematical enhancement of
longitudinal polarization.

\noindent
To make the more quantitative estimates one should take into
account the factorizable contribution (involving the parameterization
of genuine twist matrix elements) and include the $k_T$-dependent
distributions to non-factorizable one.
The consistent account for all powers of $k_T$
would be equivalent to the summation of all kinematical  high twist terms.

\noindent
To summarize, in this paper we have computed both gluon and quark contributions
to the transversely polarized $\rho$-meson electroproduction
up to genuine twist-$3$ accuracy.
We observed a number of interesting cancellations of leading infrared
divergencies due to the gauge invariance manifested through
QCD equations of motions and suggested the possible ways of treatment
of the surviving infrared contributions.

\section{Acknowledgements}

\noindent
We are grateful to
M.Diehl, A.V. Efremov, N. Kivel, B. Pire, A. Radyushkin, A. Sch{\"a}fer, C. Weiss
and M.~Vanderhaeghen for useful discussions and comments. We are also indebted
to colleagues from the University of Regensburg
(Germany), where the part of this work was performed, for warm hospitality.
This work has been supported in part by INTAS (Project 587, call 2000).

\newpage
\begin{figure}
\epsfxsize=12cm
\epsfysize=8cm
\centerline{\epsfbox{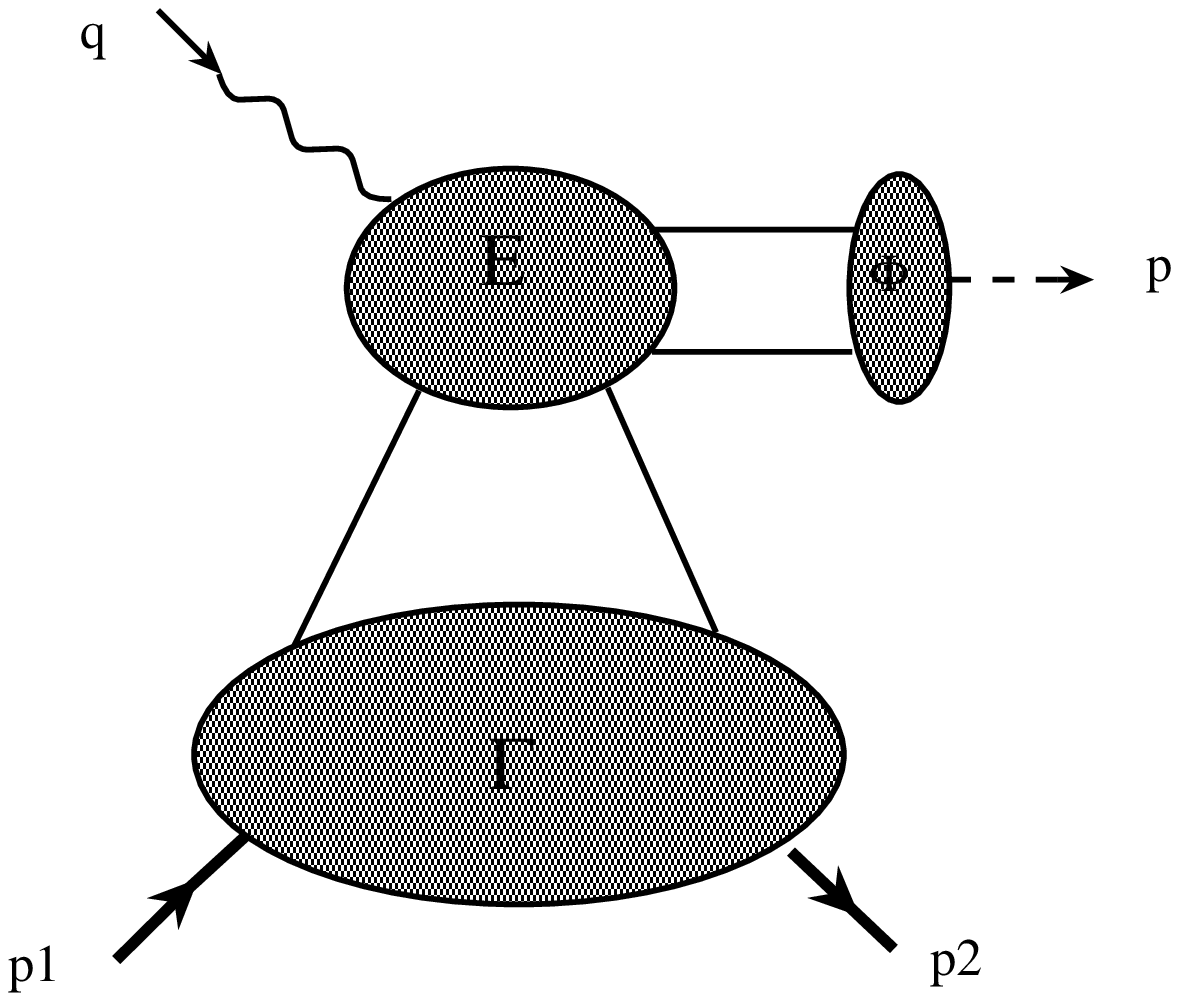}}
\caption{General structure of factorized electroproduction amplitude}
\end{figure}
\vspace{2cm}
\begin{figure}
\epsfxsize=12cm
\epsfysize=8cm
\centerline{\epsfbox{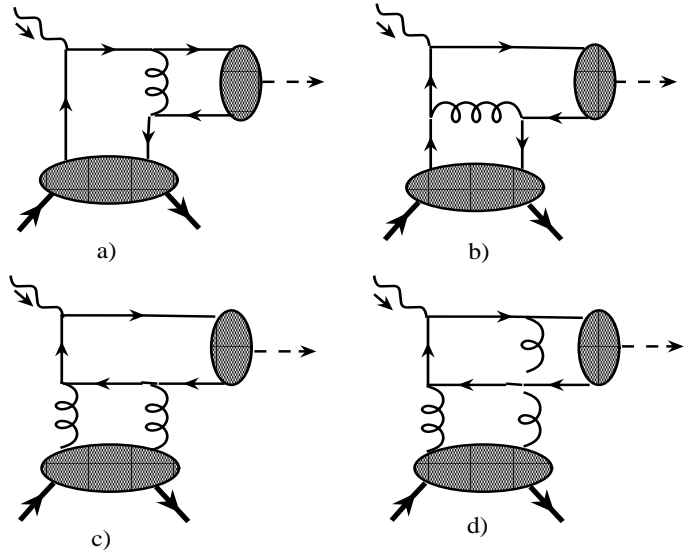}}
\caption{Simplest diagrams with quark and gluon GPD}
\end{figure}

\newpage
\begin{figure}
\epsfxsize=12cm
\epsfysize=8cm
\centerline{\epsfbox{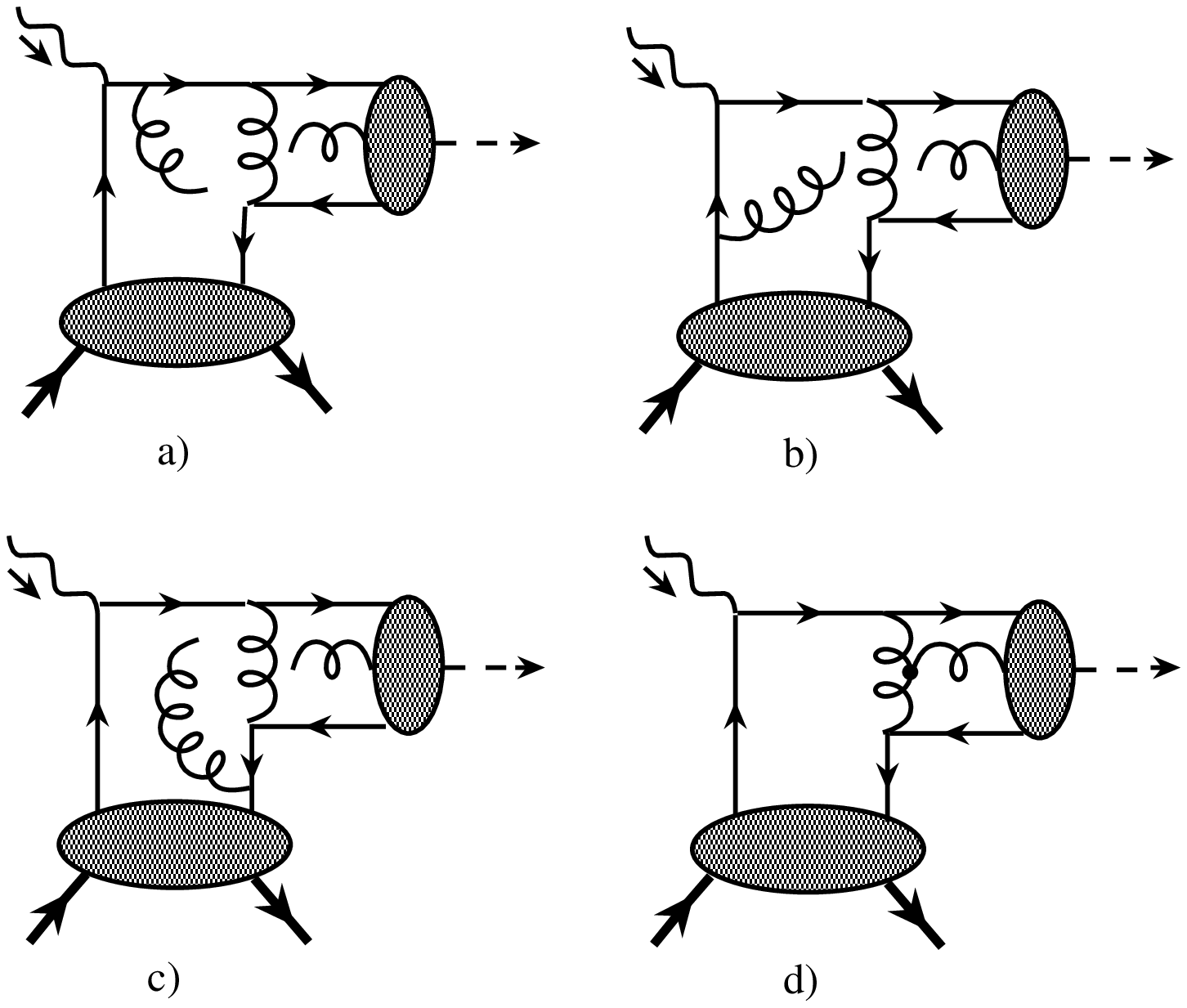}}
%\epsfxsize=12cm
%\epsfysize=8cm
%\centerline{\epsfbox{fig4_rho.ps}}
\caption{Genuine twist-$3$ diagrams with quark GPD: insertions to the
diagrams of Fig. 2a}
\end{figure}
\vspace{2cm}
\begin{figure}
\epsfxsize=12cm
\epsfysize=8cm
\centerline{\epsfbox{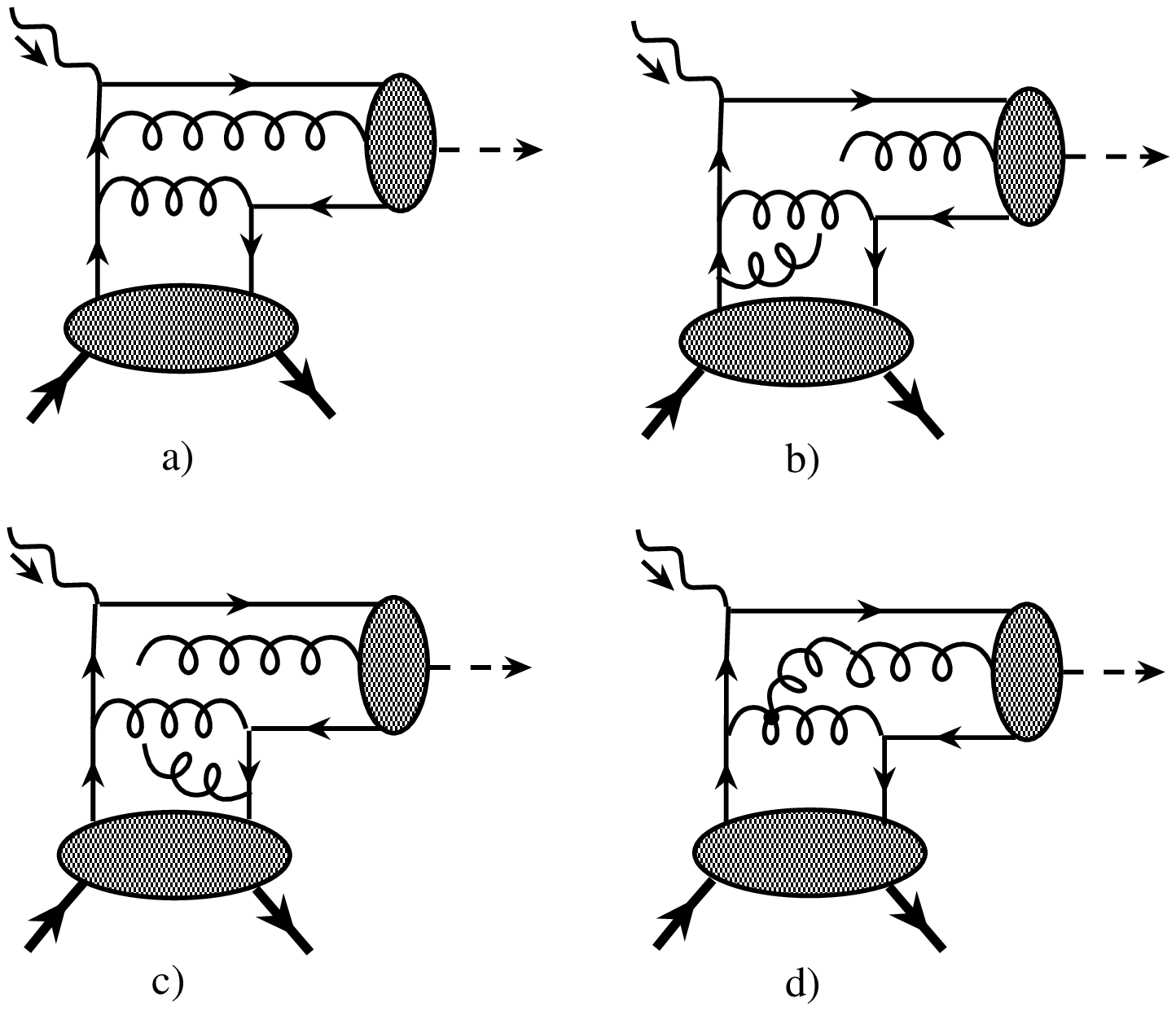}}
\caption{Genuine twist-$3$ diagrams with quark GPD: insertions
to the diagrams of Fig 2b}
\end{figure}

\newpage
\begin{figure}
\epsfxsize=12cm
\epsfysize=8cm
\centerline{\epsfbox{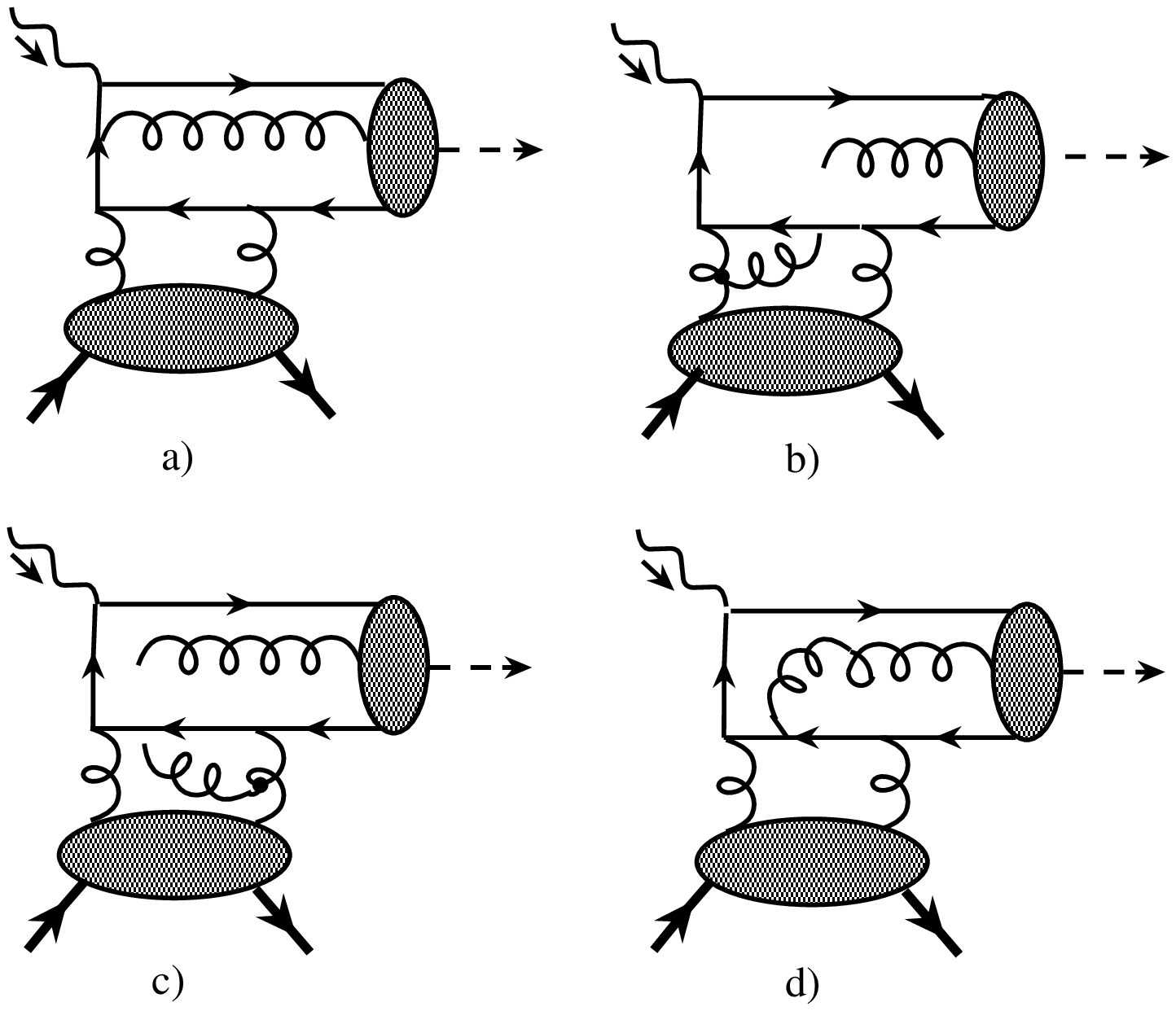}}
\caption{Genuine twist-$3$ diagrams with gluon GPD: insertions
to the diagrams of Fig 2c}
\end{figure}
\vspace{2cm}
\begin{figure}
\epsfxsize=12cm
\epsfysize=8cm
\centerline{\epsfbox{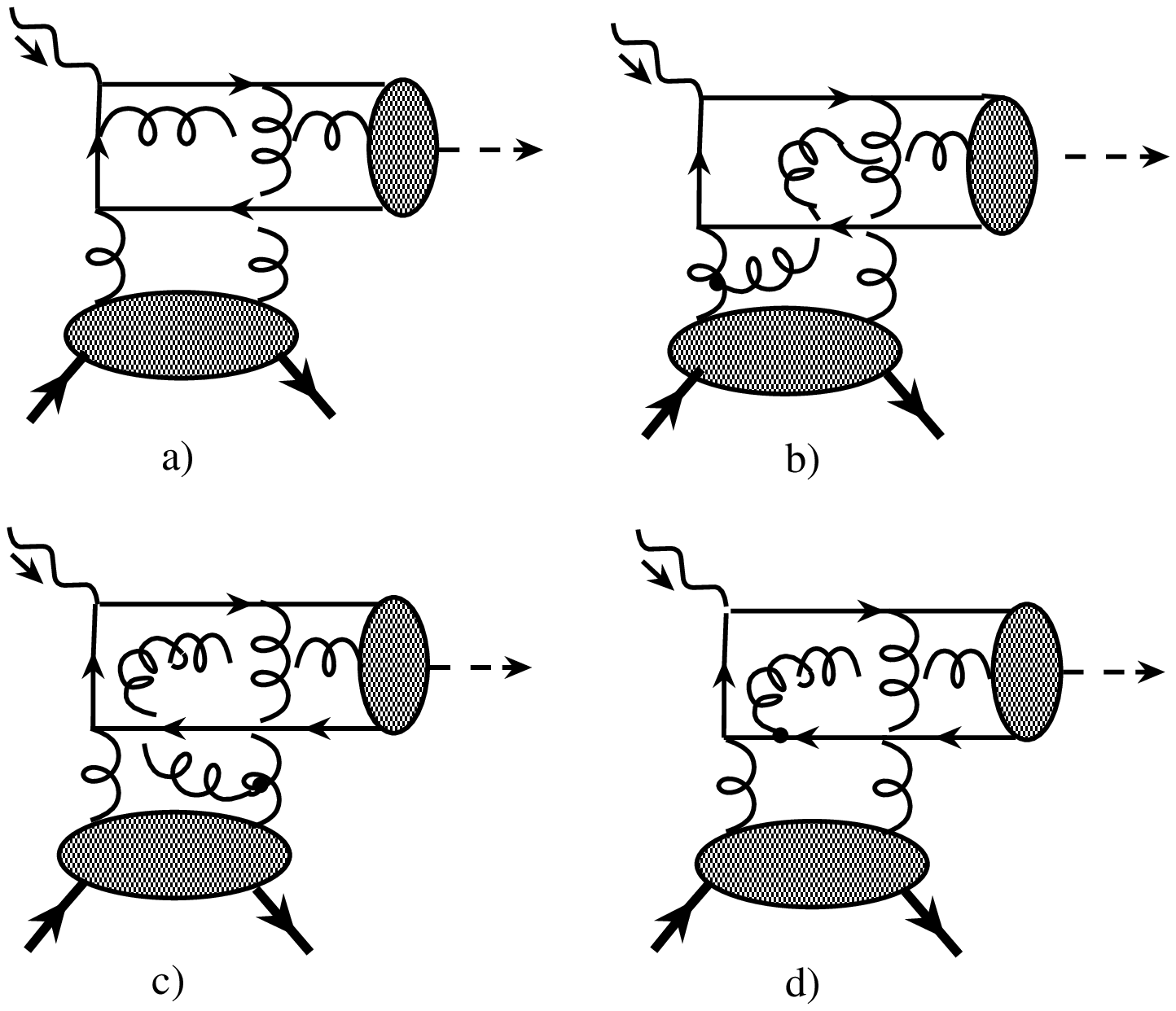}}
\caption{Genuine twist-$3$ diagrams with gluon GPD: insertions to
the diagrams of Fig 2d}
\end{figure}
\end{document}